\documentclass[conference]{IEEEtran}
\IEEEoverridecommandlockouts

\usepackage[utf8]{inputenx}

\usepackage[english]{babel}

\usepackage[final,babel=true,verbose=true]{microtype}

\usepackage{amsmath}
\usepackage{expl3}
\usepackage{calc}
\usepackage{url}
\usepackage{ifpdf}

\ifpdf
    \usepackage[pdftex,cmyk,hyperref,dvipsnames]{xcolor}
    \usepackage[pdftex]{graphicx}
\else
    \usepackage[cmyk,hyperref,dvipsnames]{xcolor}
    \usepackage{graphicx}
\fi

\usepackage[all]{foreign}

\usepackage{hyperref}
\usepackage[capitalise]{cleveref}

\usepackage[xindy,nomain,acronym,hyperfirst=false,shortcuts]{glossaries-extra}

\def\thickspace{\kern .16em}
\def\thinspace{\kern .1em}
\def\tinyspace{\kern .06em}
\def\altext#1#2{#1\thickspace/\thickspace#2}

\def\qemu{\textsc{Qemu}\xspace}

\DeclareGraphicsExtensions{.pdf,.jpeg,.png}

\hypersetup{
    plainpages=false,
    hypertexnames=true,
    linktoc=all,
    unicode=true,
    breaklinks=true,
    colorlinks=true,
    linkcolor=black,
    anchorcolor=black,
    citecolor=black,
    filecolor=black,
    menucolor=black,
    urlcolor=black
}

\setabbreviationstyle[acronym]{long-short}

\newacronym{amp}{AMP}{asymmetric multiprocessing}

\newacronym{os}{OS}{operating system}

\newacronym{ptp}{PTP}{precision time protocol}

\newacronym{rt}{RT}{real-time}
\newacronym{rtos}{RTOS}{real-time operating system}

\begin{document}

\title{
	toki: A Build- and Test-Platform for Prototyping and Evaluating Operating System Concepts in Real-Time Environments
	\thanks{
		This work was supported by the European Union (EU) under the Horizon 2020 program, projects TAPPS (Trusted Apps for open CPSs) and 5GCroCo (5G Cross Border Control), and the German Federal Ministry of Economics and Technology (BMWi) under the Smart Service World program, project PASS (Platform for Automotive Apps Guaranteeing Security and Safety).
	}
}

\author{
	\IEEEauthorblockN{Oliver Horst}%
	\IEEEauthorblockA{
		\textit{fortiss GmbH -- Research Institute of the Free State of Bavaria}\\
		Guerickestr. 25, 80805 Munich, Germany\\
		Email: horst@fortiss.org
	}
	\and
	\IEEEauthorblockN{Uwe Baumgarten}
	\IEEEauthorblockA{
		\textit{Technical University of Munich}\\
		\textit{Department of Informatics, Germany}\\
		Email: baumgaru@in.tum.de
	}
}
\maketitle

\bstctlcite{IEEE:BSTcontrol}

\begin{abstract}
	Typically, even low-level \acl{os} concepts, such as resource sharing strategies and predictability measures, are evaluated with Linux on PC hardware. This leaves a large gap to real industrial applications. Hence, the direct transfer of the results might be difficult. As a solution, we present \emph{toki}, a prototyping and evaluation platform based on FreeRTOS and several open-source libraries. toki comes with a unified build- and test-environment based on Yocto and \qemu, which makes it well suited for rapid prototyping. With its architecture chosen similar to production industrial systems, toki provides the ground work to implement early prototypes of real-time systems research results, up to technology readiness level 7, with little effort.
\end{abstract}

\section{Introduction}

Currently, most applied real-time systems research prototypes are developed and evaluated on top of Linux on PC hardware. This leaves a large gap between real industrial applications in that field and the prototype. In case of low-level \ac{os} concepts concerning, \eg context switch times, resource sharing, intra-node communication, and predictability, the drawn conclusions could even be void due to the completely different nature of the industrial platform. Furthermore, we see a lack in practical examinations of which latency and how much temporal predictability, in the sense of \cite{Sun:DoPfCPS:2016}, is achievable with certain configurations (\eg software architectures and predictability measures). Hence, we see the need to ease constructing early prototypical implementations of research results on relevant hardware in relevant environments.

On the one hand, Linux seems to be a good choice here. It is available for a wide variety of hardware platforms, provides excellent third-party library support, and can fulfill real-time requirements to some extent. However, it is a complex task to configure Linux in a way that its influence on low-level performance benchmarks is negligible or at least predictable. Moreover, integrating own concepts into the Linux kernel requires detailed knowledge about the kernel sources and its concepts. Hence, even though specialized distributions, such as LITMUS\textsuperscript{RT} \cite{calandrino_litmus^rt_2006}, can drastically reduce the configuration effort, the integration complexity remains as main issue.

Industrial-grade \acp{rtos}, on the other hand, provide considerably less influence on performance measurements, but at the costs of usability. They are either delivered as minimal systems, like FreeRTOS \cite{url:freertos}, which lack tooling support and provide not much more than a scheduler, or as sophisticated platforms tailored to specific industrial fields, such as AUTOSAR \cite{AUTOSAR:ClassicPlatformRelease:2017}, which come along with royalty fees, tooling, and complex, configurable software stacks.

As an exception, the GENODE OS framework \cite{genode} comes with toolchains for several hardware platforms and provides a convenient, production ready \ac{rtos}. Unfortunately, GENODE's overall orientation to safety and security severely hinders rapid prototyping. Its micro-kernel based design requires that all adaption and changes to the \ac{os} obey the strict isolation mantra, which is conceptually challenging and time consuming.

Therefore, we see the need for a minimal, yet flexible real-time system framework that provides comfort similar to commercial platforms, but comes without a complex software architecture and strict inherent design concepts. Ideally, the framework should be fully based on open-source software, provide an integrated standard C-library, a target toolchain, configuration tools, and an emulation environment for testing. A plus would be the possibility to safety certify the system.

Accordingly, we present \emph{toki}, a flexible and configurable \ac{os} framework, close to production industrial systems, but without the hassle of complex software architectures.

\begin{figure}[t]
    \centering
    \includegraphics[width=\linewidth]
        {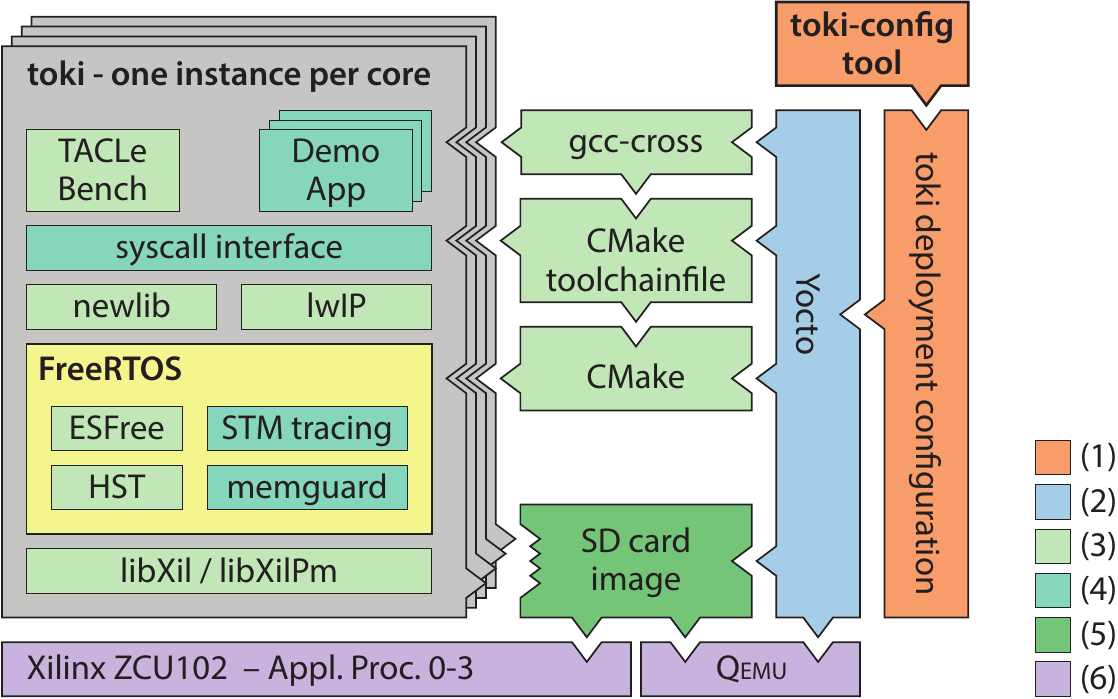}

    \caption[]{
        Sketch of toki's architecture and build-system, showing the included components (3,\thinspace4), differentiated among integrated (3) and newly created (4) components, and their relations. Given a deployment configuration (1), Yocto (2) builds all included components and collates them into a SD card image (5), which can be deployed and tested on the hardware or in \qemu (6).
    }
    \label{fig:buildflow-and-arch}
\end{figure}

\section{The Build- and Test-Platform Toki}

toki, Japanese for ``when an action occurs'', is a build- and test-environment constructed around FreeRTOS or SafeRTOS \cite{url:safertos}, respectively. toki's main goal is to compose an easy to use rapid prototyping platform to develop and evaluate new operating-system- and intra-node-communication-concepts for cyber-physical systems on commodity microcontrollers. Therefore, toki combines the following open-source projects into the unified architecture and build-system illustrated in \cref{fig:buildflow-and-arch}:
FreeRTOS \cite{url:freertos} (v10.0.0),
newlib \cite{url:newlib} (v3.0.0),
\altext{libXil}{libXilPm} \cite{url:xil-embeddedsw} (v2019.1), and
lwIP \cite{url:lwip} (v2.0.3).
These base components are supplemented by the
ESFree Scheduling Library \cite{kase_efficient_2016},
HST User Mode Scheduler \cite{paez_freertos_2015}, and
TACLe benchmarks \cite{falk_et_al:OASIcs:2016:6895}.

These components were chosen under the following aspects: \emph{(i)} the simplicity to modify and extend their code base, \emph{(ii)} their ability to closely mimic the software stacks of comparable industrial systems, \emph{(iii)} the flexibility of their design regarding rapid prototyping, and \emph{(iv)} compatible software licenses.

At platform-level, the toki build-system (\cref{fig:buildflow-and-arch}) utilizes Yocto \cite{url:yocto} to provide a self-contained build-environment that builds all components of toki together with their dependencies (\eg a GCC cross-compiler). At project-level, we ensured proper include prefixes for all components, by restructuring their sources and adding a CMake \cite{url:cmake} based build-system, where needed. This reorganization is conducted by a dedicated Python script, to allow nearly automatic upstream pulls.

To allow software developers to benefit from the code insight and debugging capabilities of their favorite IDE, single-core deployments of toki can be built by CMake. Production-ready images of multi-core deployments, on the other hand, can only be built by Yocto. toki deployment configurations, in general, are specified either manually or with assistance of the toki-config tool, and guide Yocto in the build process.

Besides the changes required for the build-environment, we contribute the following new features to FreeRTOS on ARMv8:
\acl{amp} boot support,
a newlib syscall interface,
a memguard \cite{yun_memguard:_2013} implementation, and
software-tracing support via ARM's system trace macrocell \cite{arm-stm}.

The toki build-environment is seamlessly integrated with a virtual test-environment. Hence, all images built by Yocto can either be tested on the real hardware or its emulated counterpart. The emulation is handled by a tailored \qemu \cite{url:qemu}, built by Yocto, which enables a fully virtual development cycle.

Currently, toki is solely tested and ready-to-run on the Xilinx Zynq UltraScale+ MPSoC platform \cite{url:xil-ultrascale}, a contemporary ARMv8 multi-core SoC with an integrated FPGA. This platform was selected, because of \emph{(i)} Xilinx's extensive software support for it, including a bare-metal driver kit, and \emph{(ii)} the included FPGA, which allows us to increase the accuracy of our performance evaluations. Nevertheless, toki was designed with portability in mind; hence, it can easily be ported to other target platforms, \eg the STM32Cube by STmicro \cite{url:stm32cube}.

In the future, we plan to conduct measurements to compare the interrupt handling latency of toki with other software stacks and extend toki with FreeRTOS+POSIX \cite{url:freertos+posix}, a communication middleware, and \ac{ptp} support.

Feel free to try out the latest version of toki, by downloading it from: \url{https://git.fortiss.org/toki}

\enlargethispage{1pt}

\section{Demonstration}

We will demonstrate toki through two showcases:
\begin{enumerate}
	\item
		A video of the TAPPS project's \cite{url:tapps} final demonstrator, showing the on-the-fly installation of an application into the real-time critical control path of the throttle control of a production electric motorcycle, realized with toki.
	\item
		A live-demo of toki's configuration, build, simulation, and deployment cycle on the example of memory benchmarks deployed to distinct cores and regulated by memguard.
\end{enumerate}

The first showcase focuses on the real-time capabilities and applicability to industrial use-cases, and the second on the configuration flexibility and measurement capabilities of toki.

\section*{Acknowledgment}

We would like to thank all contributors to toki for their work, namely:
Martin Jobst,
Johannes Wiesböck,
Dorel Coman,
Ulrich Huber,
Mahmoud Rushdi,
Tuan Tu Tran,
Firas Trimech,
Raphael Wild,
Andreas Ruhland, and
Dhiraj Gulati.

\bibliographystyle{IEEEtran}
\bibliography{literature}

\begin{thebibliography}{10}
\providecommand{\url}[1]{#1}
\csname url@samestyle\endcsname
\providecommand{\newblock}{\relax}
\providecommand{\bibinfo}[2]{#2}
\providecommand{\BIBentrySTDinterwordspacing}{\spaceskip=0pt\relax}
\providecommand{\BIBentryALTinterwordstretchfactor}{4}
\providecommand{\BIBentryALTinterwordspacing}{\spaceskip=\fontdimen2\font plus
\BIBentryALTinterwordstretchfactor\fontdimen3\font minus
  \fontdimen4\font\relax}
\providecommand{\BIBforeignlanguage}[2]{{%
\expandafter\ifx\csname l@#1\endcsname\relax
\typeout{** WARNING: IEEEtran.bst: No hyphenation pattern has been}%
\typeout{** loaded for the language `#1'. Using the pattern for}%
\typeout{** the default language instead.}%
\else
\language=\csname l@#1\endcsname
\fi
#2}}
\providecommand{\BIBdecl}{\relax}
\BIBdecl

\bibitem{Sun:DoPfCPS:2016}
B.~Sun \emph{et~al.}, ``Definitions of predictability for cyber physical
  systems,'' \emph{J. of Syst. Architecture - Embedded Syst. Des.}, vol.~63,
  pp. 48--60, 2016.

\bibitem{calandrino_litmus^rt_2006}
J.~M. Calandrino \emph{et~al.}, ``{LITMUS}{\textasciicircum}{RT} : {A}
  {Testbed} for {Empirically} {Comparing} {Real}-{Time} {Multiprocessor}
  {Schedulers},'' in \emph{27th {IEEE} {Int.} {Real}-{Time} {Syst.} {Symp.}
  ({RTSS})}, Dec. 2006, pp. 111--126.

\bibitem{url:freertos}
\url{https://www.freertos.org}.

\bibitem{AUTOSAR:ClassicPlatformRelease:2017}
AUTOSAR, ``Classic platform release 4.3.1,'' AUTOSAR Std., Dec. 2017.

\bibitem{genode}
N.~Feske, \emph{{GENODE} Foundations -- Operating System Framework
  19.05}.\hskip 1em plus 0.5em minus 0.4em\relax {GENODE} Labs, 2019.

\bibitem{url:safertos}
\url{https://www.highintegritysystems.com/safertos/}.

\bibitem{url:newlib}
\url{https://sourceware.org/newlib/}.

\bibitem{url:xil-embeddedsw}
\url{https://github.com/Xilinx/embeddedsw}.

\bibitem{url:lwip}
\url{https://savannah.nongnu.org/projects/lwip/}.

\bibitem{kase_efficient_2016}
R.~Kase, ``\BIBforeignlanguage{eng}{Efficient {Scheduling} {Library} for
  {FreeRTOS}},'' Master’s thesis, KTH Information and Communication
  Technology, 2016.

\bibitem{paez_freertos_2015}
F.~E. Páez \emph{et~al.}, ``{FreeRTOS} user mode scheduler for mixed critical
  systems,'' in \emph{{6th} {Argentine} {Conf.} on {Embedded} {Syst.}
  ({CASE})}, Aug. 2015.

\bibitem{falk_et_al:OASIcs:2016:6895}
H.~Falk \emph{et~al.}, ``{TACLeBench: A Benchmark Collection to Support
  Worst-Case Execution Time Research},'' in \emph{16th Int. Workshop on
  Worst-Case Execution Time Analysis ({WCET})}, 2016, pp. 2:1--2:10.

\bibitem{url:yocto}
\url{https://www.yoctoproject.org}.

\bibitem{url:cmake}
\url{https://cmake.org}.

\bibitem{yun_memguard:_2013}
H.~Yun \emph{et~al.}, ``{MemGuard}: {Memory} bandwidth reservation system for
  efficient performance isolation in multi-core platforms,'' in \emph{19th
  {IEEE} {Real}-{Time} and {Embedded} {Technol.} and {Appl.} {Symp.} ({RTAS})},
  Apr. 2013.

\bibitem{arm-stm}
\emph{CoreSight System Trace Macrocell -- Technical Reference Manual},
  r0p1~ed., ARM Limited, Dec. 2010.

\bibitem{url:qemu}
\url{https://www.qemu.org}.

\bibitem{url:xil-ultrascale}
\url{https://www.xilinx.com/products/silicon-devices/soc/zynq-ultrascale-mpsoc.html}.

\bibitem{url:stm32cube}
\url{https://www.st.com/content/st_com/en/stm32cube-ecosystem.html}.

\bibitem{url:freertos+posix}
\url{https://www.freertos.org/FreeRTOS-Plus/FreeRTOS_Plus_POSIX/}.

\bibitem{url:tapps}
\url{http://www.tapps-project.eu}.

\end{thebibliography}

\end{document}